\documentclass[floatfix,superscriptaddress]{revtex4}
\usepackage[T1]{fontenc}
\usepackage[latin9]{inputenc}
\usepackage{color}
\usepackage{soul}
\usepackage{multirow}
\usepackage{relsize}
\usepackage{graphicx}
\usepackage{calc}
\usepackage{amsmath}
\usepackage{babel}
\usepackage{float}
\usepackage{svg}
\usepackage{booktabs}
\usepackage{comment}
\usepackage{lipsum}
\usepackage{amsmath,amssymb}
\usepackage{bm}
\usepackage{float}
\usepackage[latin9]{inputenc}
\usepackage[dvipsnames]{xcolor}
\usepackage{graphicx}
\usepackage{dcolumn}
\usepackage[normalem]{ulem}
\makeatother
\usepackage{siunitx}
\usepackage[flushleft]{threeparttable}
\usepackage{xargs}                     

\graphicspath{{figs/}}
\makeatletter

\usepackage{tikz}

\usetikzlibrary{arrows.meta,positioning,calc,fit}
\usepackage{comment}
\usepackage{tabularx}
\usetikzlibrary{arrows.meta,positioning,calc,decorations.pathmorphing,decorations.markings}
\begin{document}
\title{spectrum  phase and constraints on THz-Optical klystron}

\author{Najmeh Mirian}\email{n.mirian@hzdr.de}
\affiliation{Helmholtz-Zentrum Dresden-Rossendorf HZDR, 01328 Dresden Germany}
\date{\today}
         
\begin{abstract}
Optical klystrons provide an efficient mechanism for enhancing coherent radiation through laser-induced microbunching and dispersive amplification. In the terahertz (THz) regime and at low beam energy, however, the radiation wavelength becomes comparable to the characteristic wavelengths of longitudinal space-charge (LSC) and coherent synchrotron radiation (CSR) driven microbunching. In this work, we analyze the impact of
electron-beam microbunching on the spectral amplitude and phase of a seeded optical klystron operating in the THz regime. Using a phase-space formalism, incoherent energy modulations generated by collective effects are shown to enter the bunching spectrum as a stochastic longitudinal phase, producing local wavenumber jitter and spectral broadening. An explicit connection between the microbunching-induced energy modulation
and LSC/CSR gain is established for low-energy beams, demonstrating that the overlap between collective-effect wavelengths and the optical-klystron radiation wavelength leads to strong spectral phase distortion. These effects impose fundamental constraints on the achievable harmonic
bunching and spectral purity and stability of THz optical klystrons and must be considered in the design and optimization of next-generation low-energy THz FEL facilities. 
\end{abstract}
\maketitle
\section{Introduction}

Optical klystrons provide a compact and efficient mechanism to enhance coherent radiation
via laser-induced energy modulation followed by dispersive conversion into microbunching
\cite{Vinokurov1971,Orzechowski1986,Coisson1981,Kim1986OK}.
They are widely used to increase gain, reduce saturation length, and enable harmonic generation
in seeded FEL configurations \cite{Minehara1992,Takano1991,Colson1985,Yu2000}.
Renewed interest in optical-klystron schemes is driven in part by their favorable scaling at long
radiation wavelengths, where strong bunching can be obtained with comparatively low beam energy,
supporting compact THz sources \cite{Saldin1998,SaldinBook2000,Dattoli1993,Williams2006,mirian2025ARX}.

In the terahertz regime , corresponding to wavelengths of approximately $\lambda \sim 10$--$100~\mu$m and the radiation scale shifts from the optical domain to a regime that is comparable to intrinsic collective length scales of the electron beam.
At low beam energies, these wavelengths become comparable to the characteristic interaction scales of longitudinal space charge (LSC) and coherent synchrotron radiation (CSR) of bending magnets. As a result, collective effects are no longer perturbative but strongly influence the longitudinal phase space.
In this regime, space charge-induced energy modulations and CSR-driven density perturbations can directly seed or amplify microbunching, making beam transport and compression significantly more challenging.
These effects can strongly amplify shot-noise density fluctuations, producing broadband microbunching
instability (MBI) upstream of the optical klystron \cite{Heifets2002,HuangKim2007,Saldin2004}.
As a consequence, the electron beam entering the modulator already carries microscopic density and energy modulations on the scale of the radiation wavelength, originating from collective effects rather than from the external seed.

Microbunching instability has been studied extensively in the context of high-gain FELs, primarily for
its impact on beam quality, energy spread, and gain degradation \cite{Saldin2004,HuangKim2007,SALDIN2004355}.
More recently, it has been shown that incoherent energy modulations can introduce longitudinal phase
distortions in advanced seeding schemes, leading to spectral broadening and sideband formation
\cite{Hemsing2018,mirian2021}.
The corresponding impact on the spectral amplitude and phase in seeded \emph{THz optical klystrons},
where collective-effect wavelengths overlap the radiation wavelength, has not yet been addressed
systematically.

In this work, we demonstrate that microbunching-driven energy modulations enter the optical-klystron bunching spectrum as a stochastic longitudinal phase, producing local wavenumber jitter, spectral broadening, and a reduction of coherent bunching. We explicitly connect the microbunching-induced energy modulation to the longitudinal space-charge and coherent synchrotron radiation gain for low-energy beams, thereby identifying the scaling with beam energy, current, and dispersive strength.
While the theoretical framework presented here accounts for the combined effects of the seed field and the electron-beam microstructure on the optical-klystron radiation spectrum, the following discussion assumes a shot-to-shot stable seed phase in order to isolate phase noise originating from beam-induced microstructures.

These results are particularly relevant for DALI machine, a compact low-energy THz facility under development
\cite{DALI-CDR,Mirian2025, DALI-TDR}.
DALI plans to seed the optical klystron using radiation extracted from a THz FEL oscillator, which
intrinsically limits the available seed power.
Consequently, operation in a strongly seed-dominated regime may not always be achievable, and MBI can
become comparable to (or exceed) the seed-induced energy modulation, imposing constraints on achievable
pulse energy, spectral purity, and shot-to-shot stability.

\section{Bunching Factor in an Optical Klystron}

The optical klystron consists of an energy modulator followed by a
dispersive section and represents the simplest realization of
laser-induced coherent microbunching.
Its performance is fundamentally determined by the degree of
longitudinal microbunching imprinted on the electron beam at the
radiation wavelength \cite{Penco2017, Penco2015, Thomas2010}.
The microbunching is conveniently quantified by the complex bunching
factor, which encodes both the amplitude and the longitudinal phase
coherence of the electron density modulation.
In the low-gain regime relevant to optical klystrons, and for a short
radiator operated well below saturation, the coherent radiation emitted
in the radiator scales with the square of the modulus of the bunching
factor.
In this limit, the radiated energy per pulse follows $W \propto N_e^2 |b(k)|^2$
where $N_e$ is the number of electrons and $b(k)$ is the bunching factor
evaluated at the radiation wavenumber $k$.
While the bunching amplitude determines the achievable pulse energy,
the longitudinal phase of the bunching factor governs the spectral and
temporal properties of the emitted radiation.
Consequently, any mechanism that modifies the bunching factor-either by
reducing its amplitude or by introducing phase distortions-directly
impacts the radiated power, spectral purity, and harmonic content of the
optical-klystron output \cite{Ding2006}.

Optical klystrons are most commonly operated at the
fundamental radiation wavelength or at low harmonics, typically the
second or third harmonic, where efficient bunching can be achieved with
moderate energy modulation.
Nevertheless, the mathematical description of laser-induced bunching in
an optical klystron is formally equivalent to that of high-gain harmonic
generation (HGHG) \cite{Penco2017, Penco2015}, in the low-gain, single-radiator limit.

For this reason, and to retain a compact and general notation, we
express the bunching factor using the standard HGHG formalism, in which
the harmonic index $n$ appears explicitly.
This choice allows a unified treatment of the fundamental and harmonic
operation of the optical klystron and facilitates the extension of the
analysis to higher harmonics, without implying a specific HGHG
operational mode for the facility considered here.
However, in the following analysis, we restrict the quantitative calculations to
the fundamental harmonic ($n=1$), which corresponds to the nominal
operating mode and primary scientific goal of the DALI facility.

\subsection*{Phase-space transformation}
We formulate the interaction in terms of a longitudinal phase-space transformation to describe how energy modulations are converted into longitudinal phase and density modulations.
Let $z$ denote the longitudinal coordinate and 
\begin{equation}
p\equiv\frac{E-E_{0}}{\sigma_{E}}
\end{equation}
the normalized energy deviation, where $\sigma_{E}$ is the uncorrelated
rms energy spread. After interaction with a seed laser of wavenumber
$k_{s}=2\pi/\lambda_{s}$, the electron energy is modulated as 
\begin{equation}
p_{1}=p+A(z)\sin\!\big(k_{s}z+\psi(z)\big)+\Delta p(z),
\label{eq:energ_modul}
\end{equation}
where $A(z)$ is the (possibly slowly varying) normalized modulation
amplitude, $\psi(z)$ is the optical phase of the seed. $\Delta p(z)$ in Eq. (\ref{eq:energ_modul}) 
represents additional slow energy distortions accumulated before the
dispersive section. $\Delta p(z)$ can be expressed as the superposition of
monochromatic modulations of different amplitudes: 
\begin{equation}
\Delta p(z)=\sum_{\mu}p(k_{\mu})\,\sin\!\big(k_{\mu}z+\phi_{\mu}\big),\label{eq:ok_incoherent_energy}
\end{equation}
where $p(k_{\mu})$ are the mode amplitudes,  monochromatic energy modulations at $k_{\mu}$ wavenumber, and $\phi_{\mu}$ are
uncorrelated random phases, which can be shot to shot different. We show the calculation of the incoherent energy modulations in microbunching instability section. 

We assume the modulator undulator length is short compared to both the Rayleigh length of the seed radiation and the betatron functions $\beta_{x,y}$ of the electron beam, so that diffraction and betatron focusing effects over the interaction length are negligible. The seed wavelength $\lambda_{s}$ satisfies the resonant condition
$\lambda_{s}=\frac{\lambda_{u}}{2\gamma_{0}^{2}}\left(1+K^2/2\right),$
where $\lambda_{u}$ is the undulator period and $K$ is the undulatorstrength parameter. In resonant condition $\gamma_0=E_0/mc^2$ is the reference Lorentz factor, where $mc^2$  is the electron rest energy.
Under these conditions, the interaction produces
a sinusoidal energy modulation whose amplitude depends on the transverse overlap between the seed field and the electron beam.
Neglecting higher-order effects, the amplitude of the laser-induced energy modulation can be written as \cite{Huang2010}
\begin{equation}
\Delta\gamma_{s}(z,r)=\sqrt{\frac{P_{s}(z)}{P_{0}}}\,\frac{KL_{u}}{\gamma_{0}\sigma_{r}}\left[J_{0}\!\left(\frac{K^{2}}{4+2K^{2}}\right)-J_{1}\!\left(\frac{K^{2}}{4+2K^{2}}\right)\right]\exp\!\left(-\frac{r^{2}}{4\sigma_{r}^{2}}\right),\label{eq:seed_energy_modulation}
\end{equation}
where $P_{s}(z)$ is the peak seed power profile, $P_{0}=I_{A}mc^{2}/e\simeq8.7~\mathrm{GW}$
is the Alfven power, $I_{A}$ is the Alfven current, $\sigma_{r}$
is the rms transverse size of the seed beam, $J_{0,1}$ are Bessel
functions of the first kind, and $r$ is the transverse radial coordinate.
Equation~\eqref{eq:seed_energy_modulation} shows that the energy modulation is maximum on axis and decreases radially according to the transverse Gaussian profile of the seed laser field, reflecting the finite transverse overlap between the electron beam and the radiation mode.
In the following analysis, we focus on
the on-axis modulation amplitude and assume that transverse effects can be incorporated into an effective modulation strength.
Thus $A(z)$, the normalized modulation amplitude in equation (\ref{eq:energ_modul}) is obtained as $A(z)=\Delta\gamma_{s}(z,0)mc^2/\sigma_{\sigma_E}$.

Passing through a chicane with longitudinal dispersion $R_{56}$,
the longitudinal coordinate transforms as 
\begin{equation}
z_{1}=z+R_{56}\frac{\sigma_{E}}{E_{0}}\,p_{1}.
\end{equation}

It is convenient to define the normalized dispersion parameter 
\begin{equation}
B\equiv k_{s}R_{56}\frac{\sigma_{E}}{E_{0}}.
\end{equation}

\subsection*{Bunching spectrum}

The bunching factor at wavenumber $k$ is defined as the Fourier
component of the longitudinal density, 
\begin{equation}
b(k)=\left\langle e^{-ikz_{1}}\right\rangle .
\end{equation}
Expanding the phase term using the Jacobi--Anger identity and averaging
over the initial Gaussian energy distribution yields the bunching
spectrum near the $n$th harmonic, $k\simeq k_{n}=nk_{s}$, as 
\begin{equation}
\boxed{b_{n}(k)=e^{-(nB)^{2}/2}\int dz\;f(z)\,J_{n}\!\big(-nA(z)B\big)\,\exp\!\left[-iz\big(k-k_{n}\big)+i\phi_{\mathrm{OK}}(n,z)\right],}\label{eq:ok_bunching_general}
\end{equation}
where $f(z)$ is the longitudinal current profile and $J_{n}$ is
the Bessel function of the first kind.

The effective longitudinal phase is 
\begin{equation}
\phi_{\mathrm{OK}}(n,z)=-\,nB\,\Delta p(z)+n\,\psi(z).\label{eq:ok_phase}
\end{equation}

Equation~\eqref{eq:ok_bunching_general} has the same mathematical
structure as the general EEHG bunching spectrum, but reduces to a
single Bessel factor due to the presence of only one modulator and
one dispersive section (see eq.5 in ref. \cite{mirian2021}). The
bunching factor defined in Eq.~\eqref{eq:ok_bunching_general} provides a complete description
of the spectral content of the longitudinal density modulation generated
by the optical klystron. The bunching spectrum can be interpreted
as a transform-limited (TL) envelope multiplied by a longitudinal
phase factor $\exp[i\phi_{\mathrm{OK}}(n,z)]$. It means that, in the absence
of collective effects, the bunching spectrum near the $n$th harmonic
$k\simeq k_{n}=nk_{s}$ is determined by the seed-induced energy modulation
amplitude, the dispersive strength, and the longitudinal envelope
and phase of the seed laser $\psi(z)$, yielding a transform-limited
spectral distribution centered at $k_{n}$. When incoherent energy
modulations are present upstream of the dispersive section, however,
the bunching spectrum acquires an additional longitudinal phase $-nB\,\Delta p(z)$. 

A useful physical interpretation is obtained by introducing the instantaneous
longitudinal bunching wavenumber, 
\begin{equation}
k_{z}=k_{n}+\phi'_{\mathrm{OK}}(n,z).
\end{equation}
this gives us
\begin{equation}
\boxed{
k_{z}=k_{n}+\psi'-nB\sum_{\mu}p(k_{\mu})k_{\mu}\cos\!\left(k_{\mu}z+\phi_{\mu}\right).}
\label{eq:kvalue}
\end{equation}
Thus, incoherent energy modulations generate stochastic fluctuations
of the local bunching wavenumber around $k_{n}$.
When the characteristic wavelengths of these fluctuations are long
compared to the bunching envelope, the phase derivative varies slowly
along the seeded region and acts primarily as an effective energy
chirp, leading to a deterministic shift of the spectral centroid and
a moderate increase of the bandwidth. In contrast, energy modulations
with wavelengths comparable to the radiation scale ($k_{\mu}\sim k_{n}$)
induce rapid, shot-to-shot fluctuations of $k_{z}$. 

We define the rms spectral bandwidth of $|b_{n}(k)|^{2}$ as 
\begin{equation}
\sigma_{k}^{2}\;\equiv\;\langle(k-\langle k\rangle)^{2}\rangle\;\simeq\;\sigma_{k,\mathrm{TL}}^{2}+\sigma_{\phi'_{\mathrm{OK}}}^{2},\label{eq:sigk_def}
\end{equation}
where $\sigma_{k,\mathrm{TL}}$ is the transform-limited bandwidth
set by the bunching envelope (i.e., the longitudinal profiles $f(z)$
and $J_n(-nA(z)B)$ ), and 
\[
\sigma_{\phi'_{\mathrm{OK}}}^{2}\equiv\langle[\phi'_{\mathrm{OK}}(n,z)-\langle\phi'_{\mathrm{OK}}(n,z)\rangle]^{2}\rangle,\]
is the excess bandwidth due to seed phase and electron sub structure
phase. Assuming (i) the microbunching wavelengths are short compared
to the bunching envelope so that many oscillations occur within the
seeded region, and (ii) the phases $\phi_{\mu}$ are uncorrelated
over $\mu$, the variance becomes 
\begin{equation}
\boxed{
\sigma_{\phi'_{\mathrm{OK}}}^{2}\;\simeq\;n^{2}\,\sigma_{\psi'}^{2}\;+\;\frac{(nB)^{2}}{2}\sum_{\mu}\big[p(k_{\mu})\,k_{\mu}\big]^{2},
}
\label{eq:sigk_OK_analog}
\end{equation}
where $\sigma_{\psi'}^{2}$ accounts for any residual seed-phase chirp
or phase noise (if present), and the second term is the microbunching-induced
contribution. In the common case of a shot-to-shot stable seed with
constant phase, $\sigma_{\psi'}=0$ and 
\begin{equation}
\sigma_{k}^{2}\;\simeq\;\sigma_{k,\mathrm{TL}}^{2}\;+\;\frac{(nB)^{2}}{2}\sum_{\mu}\big[p(k_{\mu})\,k_{\mu}\big]^{2}.\label{eq:sigk_OK_simple}
\end{equation}
Equation~(\ref{eq:sigk_OK_simple}) shows broadband microbunching
imprints a stochastic longitudinal phase whose derivative produces
wavenumber jitter and therefore an additive growth of the rms spectral
bandwidth. Because the coefficient scales as $(nB)^{2}$, harmonic
operation can be particularly sensitive even to
modest incoherent energy modulations.

To isolate the spectral distortions originating from electron-beam
microbunching, we consider in the following a constant (or shot-to-shot
stable) seed phase, $\psi(z)=\psi_{0}$, which contributes only an
overall constant phase factor $e^{in\psi_{0}}$ and therefore does
not affect $|b_{n}(k)|^{2}$. Under this assumption, all spectral
distortions discussed below originate from the microbunching-induced
energy modulation $\Delta p(z)$, which acts as a stochastic longitudinal phase imprint
on the density modulation.

In the absence of collective effects, the longitudinal phase of the
bunching factor remains constant across the spectral bandwidth, yielding
a transform-limited bunching spectrum centered at the seed wavenumber.
When collective effects become significant, this condition is violated:
microbunching-induced energy modulations introduce stochastic
longitudinal phase variations, leading to spectral broadening,
pedestals, sidebands and phase instabilty in the bunching spectrum.

This effect is particularly pronounced in the THz regime. At low beam
energy, longitudinal space charge and coherent synchrotron radiation
efficiently amplify microscopic density fluctuations at wavelengths
overlapping the optical-klystron radiation. Because the normalized
dispersion $B$ must be large to achieve significant harmonic bunching,
even modest incoherent energy modulations are converted into sizable
longitudinal phase excursions. As a result, the optical-klystron bunching
spectrum is no longer determined solely by the seed laser and the
dispersive section, but is fundamentally constrained by collective
effects in the electron beam.

The bunching spectrum therefore provides a direct and sensitive diagnostic
of microbunching instability in THz optical klystrons, linking the
spectral purity and stability of the emitted radiation to the broadband
energy modulation generated by LSC and CSR upstream of the dispersive
section. 

\subsection*{Factorized form}

If the modulation amplitude $A(z)$ varies slowly on the scale of
the optical wavelength, the Bessel function may be factorized as 
\begin{equation}
J_{n}\!\big(-nA(z)B\big)\simeq g(z)\,J_{n}(-nAB),
\end{equation}
where $A$ is the peak modulation amplitude and $g(z)$ describes
the longitudinal envelope. In this case the bunching spectrum becomes
\begin{equation}
b_{n}(k)\simeq\bar{b}_{n}\int dz\;f(z)\,g(z)\,\exp\!\left[-iz\big(k-k_{n}\big)+i\phi_{\mathrm{OK}}(z)\right],
\label{eq:ok_bunching_microbunching}
\end{equation}
with the intrinsic optical-klystron bunching amplitude 
\begin{equation}
\boxed{\bar{b}_{n}=e^{-(nB)^{2}/2}\,J_{n}(-nAB).}\label{eq:ok_bunching_factor}
\end{equation}

This expression shows explicitly the competition between laser-induced energy modulation
and the suppression of bunching due to uncorrelated energy spread in the dispersive section.

At the central harmonic $k=k_{n}$, incoherent energy modulations
also reduce the coherent bunching amplitude. Expanding the phase term
in Eq.~\eqref{eq:ok_bunching_microbunching} yields 
\begin{equation}
b_{n}(k_{n})\approx\bar{b}_{n}\prod_{\mu}J_{0}\!\big(nB\,p(k_{\mu})\big),\label{eq:ok_amplitude_product}
\end{equation}
where $\tilde{b}_{n}$ is the bunching amplitude in the absence of
microbunching. For weak incoherent modulations, $|nB\,p(k_{\mu})|\ll1$,
this reduces to 
\begin{equation}
\boxed{
b_{n}(k_{n})\approx\bar{b}_{n}\prod_{\mu}\left(1-\frac{(nB\,p(k_{\mu}))^{2}}{4}\right),}
\label{eq:ok_bunching_factorMBI}
\end{equation}
demonstrating that increasing microbunching strength leads to a monotonic
reduction of the coherent optical-klystron bunching peak.

\section{Origin of the incoherent energy modulations}
\label{sec:MBI_OK}

To evaluate the impact of microbunching instability on the optical-klystron
bunching spectrum, we model the broadband incoherent energy modulations that develop
upstream of, and within, the modulator due to collective effects, primarily LSC and CSR in bunch-compressor bending
sections. We represent the accumulated energy distortion at the modulator exit as a
superposition of sinusoidal components in Eq.~(\ref{eq:ok_incoherent_energy}), where
$p(k_\mu)$ are the (normalized) broadband energy-modulation amplitudes and $\phi_\mu$
are random phases. 

In the linear regime, the MBI-driven energy modulation amplitude at wavenumber $k_\mu$
accumulated across a modulator of length $L_u$ can be written as
\cite{Perosa2021,DiMitri2025,Huang2004,Huang2010}
\begin{equation}
p(k_\mu)=4\pi\, b_\mu(k_\mu)\,
\frac{I}{I_A}\,
\frac{Z_{\mathrm{LSC}}(k_\mu)}{Z_0\,\sigma_\gamma}\,L_u,
\label{eq:energyMBI}
\end{equation}
where $b_\mu(k_\mu)$ is the broadband density bunching factor at the modulator entrance
(shot-noise seeded in the absence of an imposed modulation), $I$ is the peak current,
$I_A=17.045~\mathrm{kA}$ is the Alfv\'en current, $Z_0=377~\Omega$ is the free-space impedance,
and $\sigma_\gamma=\sigma_E/(mc^2)$ is the uncorrelated rms slice energy spread normalized
to the rest energy.

For an in-vacuum beam, the LSC impedance per unit length is approximated by
\cite{Huang2004,Huang2010}
\begin{equation}
Z_{\mathrm{LSC}}(k_\mu)=
i\,\frac{Z_0 k_\mu}{4\pi\gamma_z^2}
\left[1 + 2\ln\!\left(\frac{\gamma_z}{k_\mu r_b}\right) \right],
\label{eq:Z_LSC_OK}
\end{equation}
where $\gamma_z=\gamma/\sqrt{1+K_u^2/2}$ is the longitudinal Lorentz factor inside an
undulator with the (peak) undulator parameter $K_u$, and $r_b$ is the effective transverse
beam size. We use $r_b=0.8735\sqrt{\epsilon_x\beta_x+\epsilon_y\beta_y}$, where $\epsilon_{x,y}$
and $\beta_{x,y}$ are the rms emittances and betatron functions in the transverse planes.

The MBI-induced energy modulations in Eq.~(\ref{eq:energyMBI}) are evaluated numerically
using a comprehensive linear gain model of the instability, tracking its evolution from
beam injection through the accelerator up to the undulator line. The model includes the
effects of longitudinal energy dispersion ($R_{56}$) and coherent synchrotron radiation
(CSR) in the magnetic bunch compressor, as well as longitudinal space charge (LSC) at low
beam energy~\cite{DiMitri2020, Perosa2021}. Starting from a shot-noise-like initial bunching
factor, the model provides the evolution of the broadband bunching spectrum at arbitrary
locations along the beamline.

In the simplified case of a single-stage bunch compression and under the assumption of
linear instability gain, the amplification of a density modulation (microbunching gain)
\cite{Huang2002,Stupakov2002}, in the presence of an arbitrary incoming energy distribution
$V(P_{0})$~\cite{Huang2004}, can be expressed as
\begin{eqnarray}
G & = & \left|\frac{b_{\mu f}}{b_{\mu0}}\right|
\backsimeq
\frac{I}{\gamma I_{A}}
\left|k_{\mu f}R_{56}\int_{0}^{L}ds\,\frac{4\pi Z(k_{\mu0},s)}{Z_{0}}\right|
\int dP_{0}\,V(P_{0})\,e^{-ik_{\mu f}R_{56}P_{0}},
\label{eq:MBIgain}
\end{eqnarray}
where $b_{\mu0}$ and $b_{\mu f}$ are the bunching factors at the entrance and exit of the
compressor section, respectively, $Z(k_{\mu0},s)$ is the longitudinal impedance per unit
length (including LSC and CSR contributions) evaluated along the line, and $P_0$ denotes the
normalized energy variable used in the distribution $V(P_0)$. The wavenumber transforms as
$k_{\mu f}=k_{\mu0}/(1+hR_{56})$, where $h$ is the initial linear energy chirp.

\begin{figure}
\centering \includegraphics[width=8 cm]{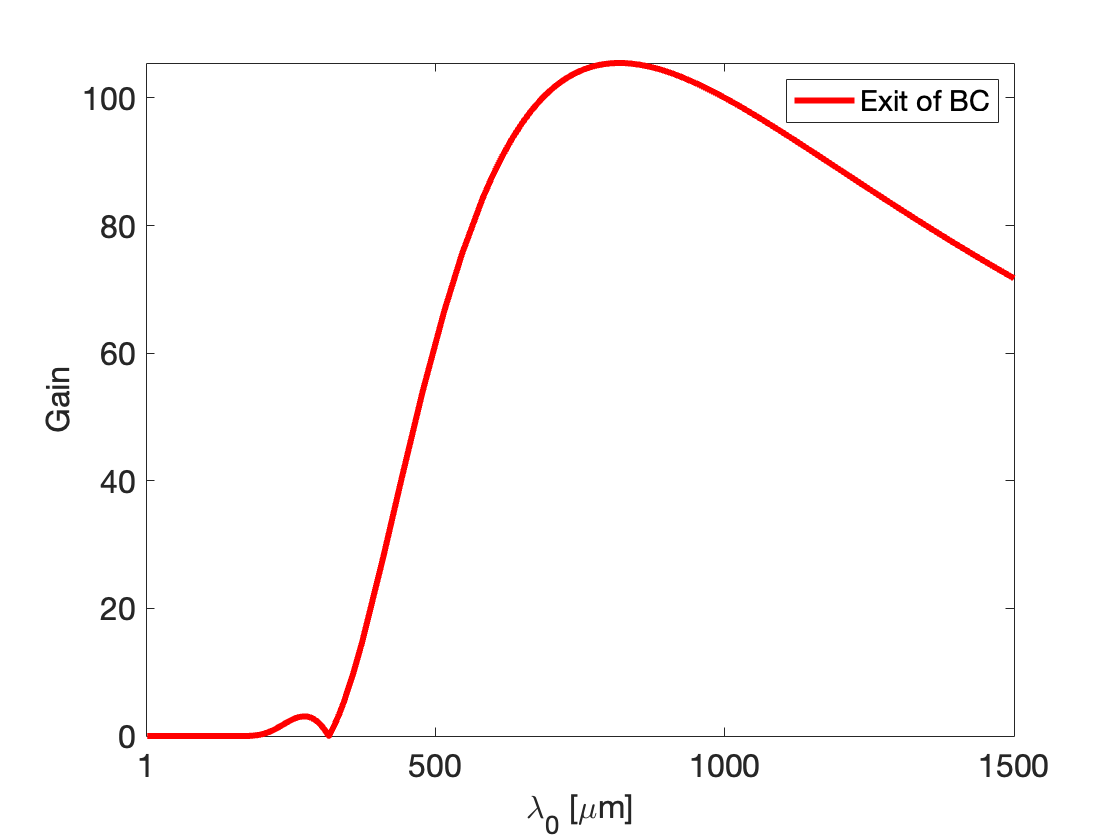} \caption{Microbunching gain at the exit of the beamline section upstream of
the dispersive element.}
\label{fig:MBI_gain} 
\end{figure}
\section{simulation and results with DALI beam parameters and operating point}
\subsection*{Microbunching instability calculation}
The analysis presented above is now applied to the parameter regime of the DALI facility. In the operating point considered here, the optical
klystron performance is defined by a compressed electron beam with peak current $I \simeq 1~\mathrm{kA}$ and rms bunch length $\sigma_z \simeq 124~\mu\mathrm{m}$, chosen to enable the generation of intense, short THz radiation pulses.

Using the linear microbunching gain model described in
Sec.~\ref{sec:MBI_OK}, the broadband density modulation seeded by shot noise is tracked from the injector through the accelerating sections and the magnetic bunch compressor up to the entrance of the optical klystron modulator. The resulting gain spectrum $G(k_\mu)$ reflects the combined action of longitudinal space charge and coherent synchrotron radiation in the bunch-compressor bending sections under DALI operating conditions.

The electron beam parameters considered in this study-50 MeV beam energy, 1 nC bunch charge, and an rms bunch length of $124~\mu\mathrm{m}$ match the design goals of the DALI facility and define an ambitious yet well-justified operating point for intense THz production and optical klystron schemes. Reaching these conditions at relatively low beam energies is technically challenging, mainly because collective effects such as space charge and coherent synchrotron radiation become significant. Nevertheless, comparable parameter ranges have been demonstrated at several accelerator facilities, confirming the feasibility of producing high-charge (hundreds of pC up to the nC level) and sub-picosecond electron bunches in the tens-of-MeV regime, even if not always achieved simultaneously within a single operational setup \cite{Zhang2017}. The layout and beam dynamics of DALI machine is under the investigation and will be reported in near future. 

The main linac section considered for the DALI facility incorporates a superconducting RF (SRF) gun optimized for high-charge, high-repetition-rate operation. It is intended to deliver electron bunches of up to 1 nC at a repetition rate of 1 MHz to drive superradiant and optical klystron THz free-electron laser (FEL) beamlines.
In the present study, we consider a representative operating scenario in which an electron bunch with an initial peak current of approximately 
100 A is accelerated to a beam energy of 50 MeV using four superconducting TESLA-type RF cavities (each approximately 1.25 m in length). A 4 m-long diagnostic section is assumed between the second and third cavities (2+2 configuration). Downstream of the linac, the beam is compressed with an assumed compression factor of 10.
Since the detailed and finalized DALI lattice is still under development, the microbunching analysis presented here is performed for an idealized and approximately optimized configuration. The microbunching gain shown in Fig.~\ref{fig:MBI_gain} therefore represents an indicative estimate under near-optimal compression conditions, rather than a prediction for a finalized machine layout. The calculation is intended to evaluate the order of magnitude of the instability and to identify the relevant parameter sensitivities.
 
Figure \ref{fig:MBI_gain} shows the calculated microbunching gain after bunch compressor respect to the uncompressed microbunching wavelength. 

\begin{figure}
\centering \includegraphics[width=8 cm]{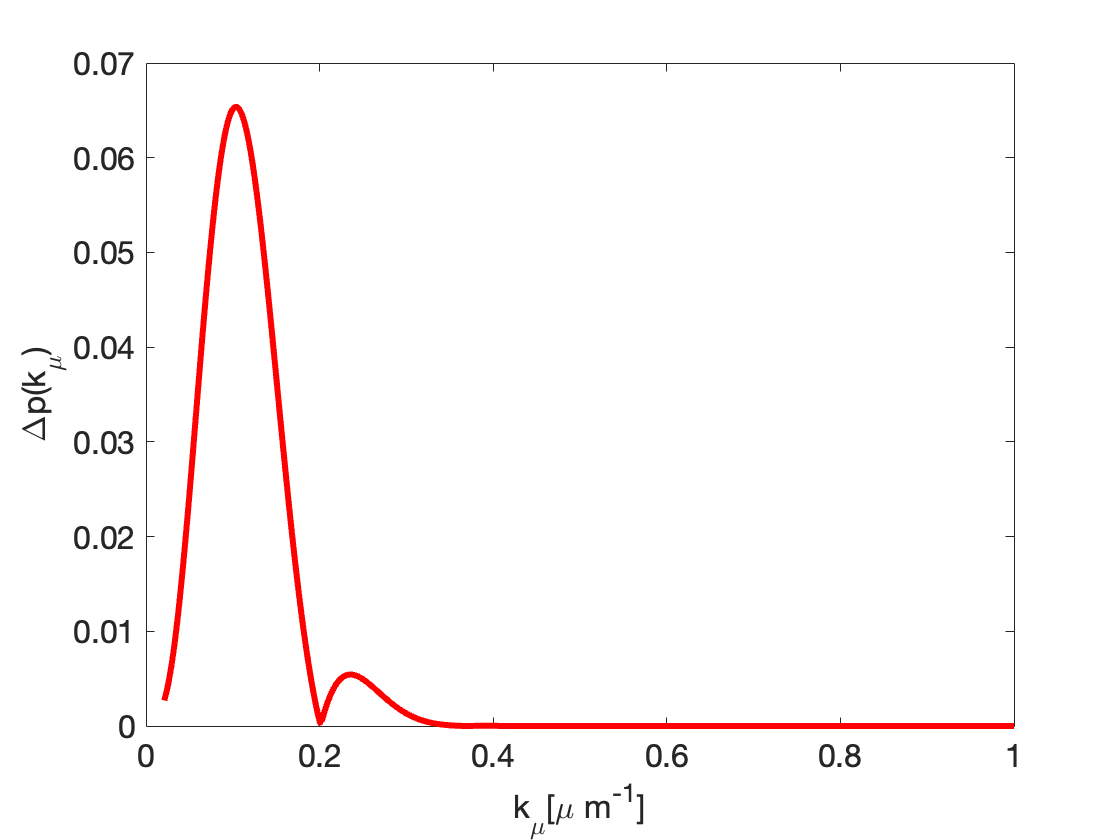} \caption{The calculated relative energy modulation $p(k_{\mu})$ due to microbunching instability at the modulator exit }
\label{fig:MBI_Pmu} 
\end{figure}

The amplified density modulations at the modulator entrance are subsequently converted into incoherent energy modulation along the modulator according to Eq.~(\ref{eq:energyMBI}). Figure \ref{fig:MBI_Pmu} demonstrates the relative energy modulation $p(k_mu)$ at the modulator exit exhibits. For this calculation, the undulator is assumed to be opened, K=0.  
For the DALI parameter set, the accumulated energy modulation spectrum $p(k_\mu)$ at the modulator exit exhibits significant amplitude in the wavelength range overlapping the optical-klystron radiation.

\subsection*{Seed laser interaction with the electron beam in the modulator}
\label{subsec:seed_modulator}
\begin{figure}
\centering \includegraphics[width=8 cm]{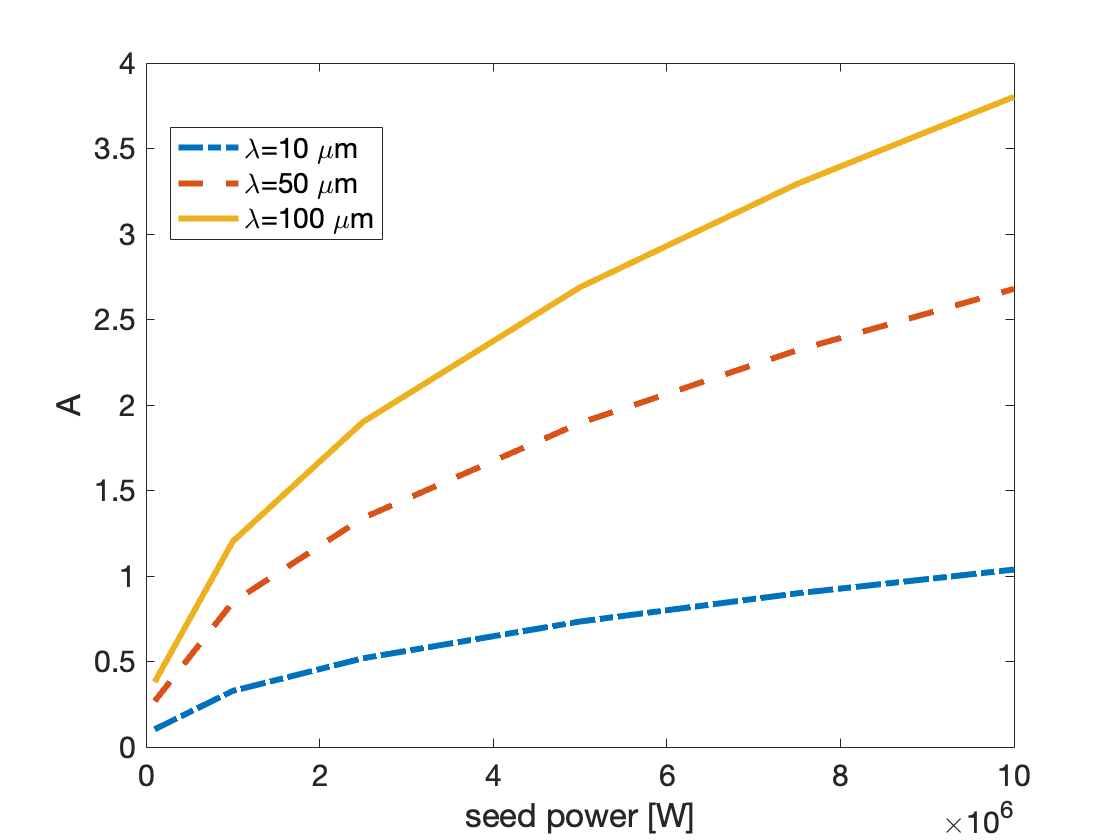} \caption{Normalized
seed-induced energy modulation amplitude $A=\Delta E/E$ as a function of the seed laser power for three seed wavelengths,
$\lambda = 100~\mu$m, $50~\mu$m, and $10~\mu$m, calculated using the DALI beam parameters in eq.(\ref{eq:seed_energy_modulation}).
Longer seed wavelengths lead to a stronger energy modulation at a given seed power. Calculations are performed for a beam energy
$E_{0}=50$ MeV ($\gamma_{0}\simeq98.9$), rms energy spread $\sigma_{E}=100$
keV, undulator length $L_{u}=1$ m with $N_{u}=10$ periods,  and rms transverse seed beam size $\sigma_{r}=2$
mm.}%
\label{fig:A} 
\end{figure}

In the proposed DALI optical klystron, the electron beam is energy-modulated by an external radiation field provided by a FEL oscillator operating at THz
wavelengths with second LINAC. The oscillator radiation is injected into the modulator undulator and serves as the seed for the optical-klystron interaction.
Because the seed is generated by the oscillator, its peak
power is intrinsically limited.
To account for this constraint, the optical-klystron performance is
evaluated over a representative range of seed peak powers, spanning
from $100~\mathrm{kW}$ to $80~\mathrm{MW}$.  
We consider a fundamental Gaussian radiation mode co-propagating with a round electron beam through a planar undulator of length $L_{u}=1$. The seed pulse is assumed to be much longer than the electron bunch, so that its amplitude and phase can be treated as constant over the interaction region ($\psi(z)= Constance$ in Eq. \ref{eq:energ_modul}). Applying the DALI beam parameters to Eq.~(\ref{eq:seed_energy_modulation}), we calculate the relative energy modulation amplitude $A$, plotted in Fig.~\ref{fig:A}. Figure~\ref{fig:A} shows the relative energy modulation amplitude $A$ induced in the electron beam as a function
of the seed laser power for different seed wavelengths.
For a fixed seed power, the modulation amplitude decreases significantly as the seed wavelength decreases.
This behavior reflects the weaker longitudinal electric field associated with shorter-wavelength radiation,
which reduce the beam-laser interaction in the modulator.
As a result, substantially stronger seed power is required at shorter wavelengths to achieve a given level of
energy modulation.

In order to optimise the bunching factor, the chicane compaction $R_{56}$ is
scanned around the theoretical optimum inferred from the seed-induced energy
modulation amplitude $A=\Delta E/\sigma_E$ appearing in the Bessel function of
Eq.~(\ref{eq:ok_bunching_factor}). 
Maximisation of the fundamental bunching in the limit of weak energy-spread
smearing yields the condition $B = 1.84/A$, corresponding to the first maximum
of the Bessel function $J_1$. 
This leads to the approximate optimum
\begin{equation}
R_{56}^{\mathrm{opt}} \approx \frac{1.84}{k_s}\frac{E_0}{\Delta E},
\qquad
k_s = \frac{2\pi}{\lambda_s}.
\end{equation}

\subsection{Pulse intensity reduction due to MBI}
In a seeded optical-klystron configuration, the radiation pulse energy is
directly determined by the magnitude of the coherent bunching factor at the
radiation wavenumber. In the small-signal regime, the emitted pulse intensity
scales proportionally to the square of the bunching amplitude,
$Int \propto |b(k)|^2$. 
Consequently, any degradation of the bunching factor caused by microbunching
instability (MBI) translates directly into a reduction of the output pulse
intensity.
\begin{figure}
\centering \includegraphics[width=8 cm]{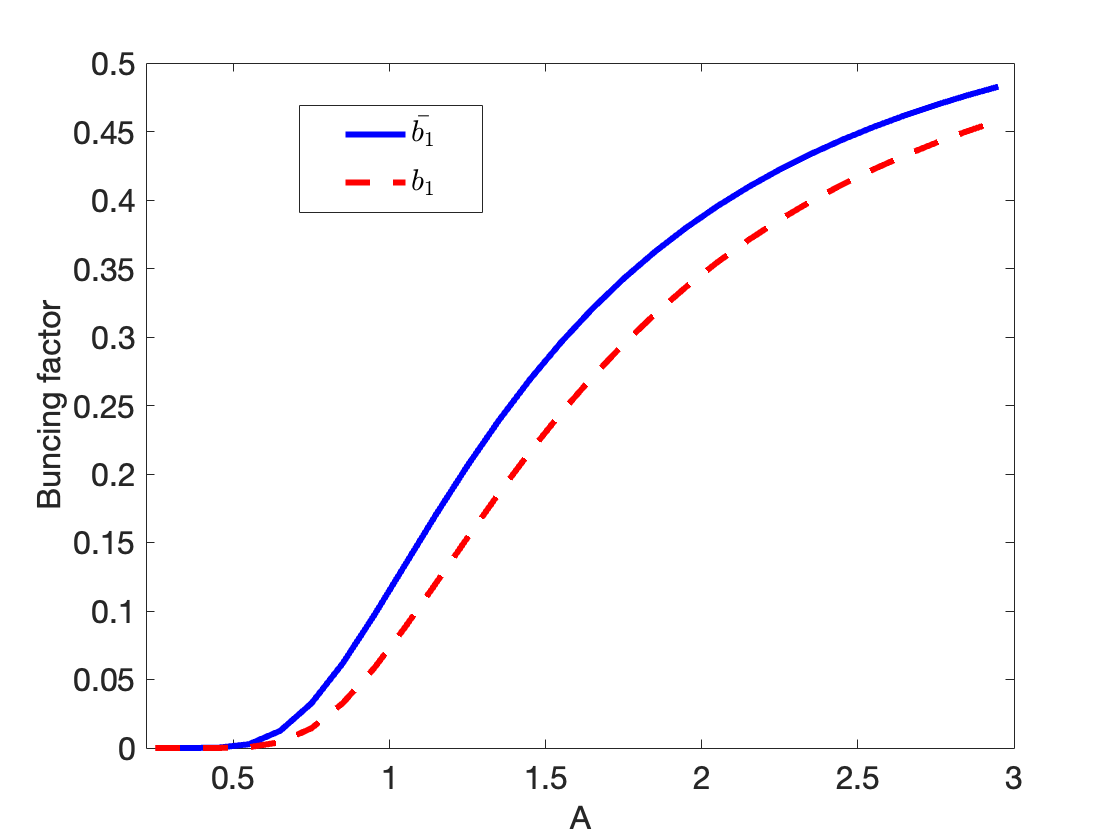} \caption{Fundamental bunching amplitude as a function of the normalized
seed-induced energy modulation $A$. The blue solid curve corresponds to the
ideal case without microbunching instability (MBI), calculated using
Eq.~(\ref{eq:ok_bunching_factor}), while the red dashed curve includes the effect of
MBI-induced energy spread according to Eq.~(\ref{eq:ok_bunching_factorMBI}).
The presence of MBI leads to a systematic reduction of the achievable bunching
amplitude over the entire parameter range. }
\label{fig:buncing_factor} 
\end{figure}
Using the beam and lattice parameters introduced in the previous subsection,
the MBI-induced energy modulations are first evaluated and propagated through
the dispersive section. The resulting bunching factor for fundamental harmonic (n=1) is then calculated using
Eqs.~(\ref{eq:ok_bunching_factor}) and (\ref{eq:ok_bunching_factorMBI}), taking into account both the seed-induced
energy modulation and the additional incoherent energy modulations by MBI.
The corresponding bunching amplitudes obtained from the data shown in
Figs. \ref{fig:MBI_Pmu} and \ref{fig:A} are summarized in Fig.~\ref{fig:buncing_factor}.

As shown in Fig.~4, the inclusion of MBI-induced energy modulation results in a
clear suppression of the fundamental bunching amplitude compared to the ideal case.
Since the emitted radiation pulse intensity scales as $|b|^2$, this reduction
directly translates into a lower THz pulse energy.
The relative impact of MBI increases with decreasing seed dominance, i.e.\ when
the MBI-induced modulation becomes comparable to the seed-induced energy
modulation.

\subsection{Bandwidth enlargement and central frequency fluctuation}

In a seeded optical-klystron configuration, the spectral bandwidth of the
emitted radiation is determined by the spectral width of the bunching factor
around the resonant wavenumber. In the absence of collective effects, the
bandwidth is primarily limited by the finite longitudinal extent of the
electron beam and laser pulse profile. In the long-seed limit relevant for the DALI configuration, the transform-limited
bandwidth is determined solely by the longitudinal extent of the electron beam, ' if we ignore the slippage effect'.
Rather than attempting to model this contribution in detail, we focus on the
additional spectral broadening induced by microbunching instability.
Accordingly, we evaluate only the excess bandwidth
$\Delta\sigma_k^2 = \sigma_k^2 - \sigma_{k,\mathrm{TL}}^2$,
given by the second term in Eq.(\ref{eq:sigk_OK_simple}), which depends explicitly on the MBI-driven bunching spectrum.
For the DALI parameter regime considered here, the seed pulse is
much longer than the electron bunch, and therefore does not contribute to the
spectral broadening. 

Using the MBI-induced energy modulation data presented in Figs. \ref{fig:MBI_Pmu} and \ref{fig:A}, the
resulting bandwidth enlargement is calculated according to Eq.~(\ref{eq:sigk_OK_simple}).
\begin{figure}
\centering \includegraphics[width=0.58\linewidth]{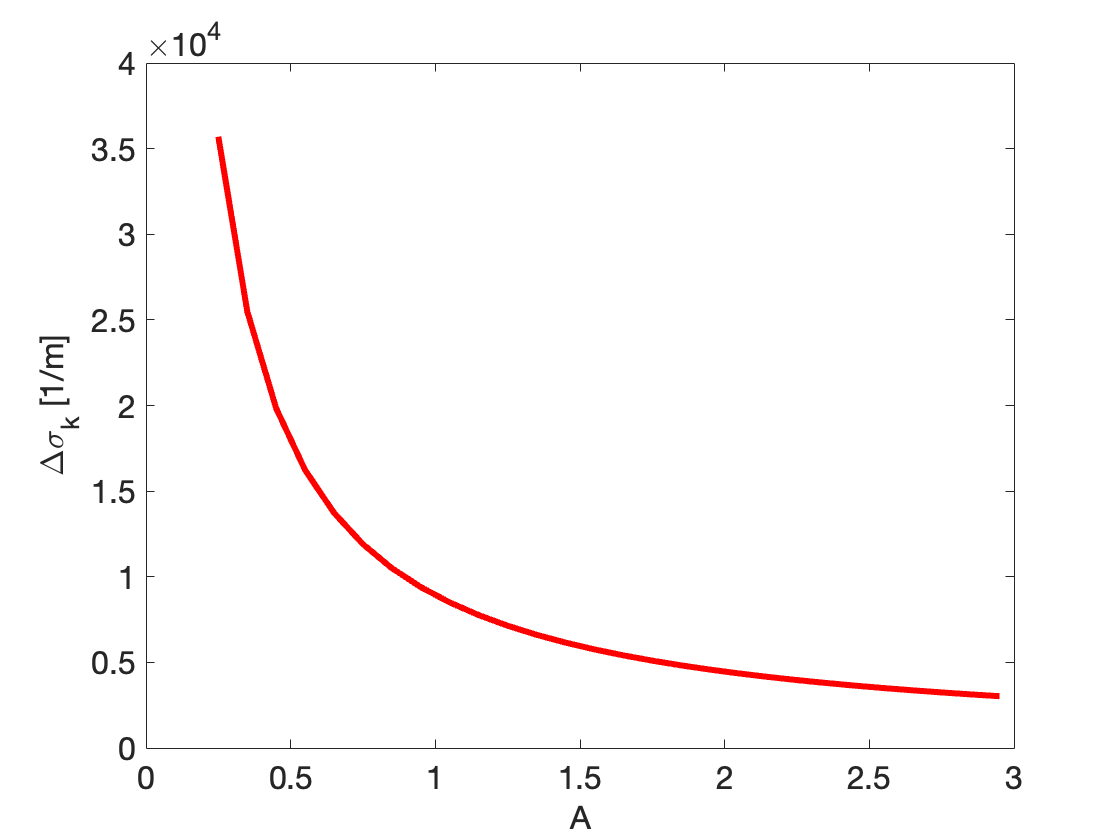} \caption{ Excess rms spectral bandwidth $\Delta\sigma_k$ induced by
microbunching instability as a function of the normalized seed-induced energy
modulation amplitude $A$.
The quantity $\Delta\sigma_k$ is obtained from the second term of
Eq.~(\ref{eq:sigk_OK_simple}) and represents the contribution beyond the
transform-limited bandwidth set by the electron bunch length.
 }
\label{fig:MBI_sigma} 
\end{figure}
Figure~\ref{fig:MBI_sigma} shows the MBI-induced excess rms bandwidth
$\Delta\sigma_k = \sqrt{\sigma_k^2 - \sigma_{k,\mathrm{TL}}^2}$ as a function of
the normalized seed modulation amplitude $A$.
The excess bandwidth decreases monotonically with increasing $A$, reflecting
the reduced relative impact of MBI as the seed-induced energy modulation becomes
dominant.
\begin{figure}
\centering \includegraphics[width=0.6\linewidth]{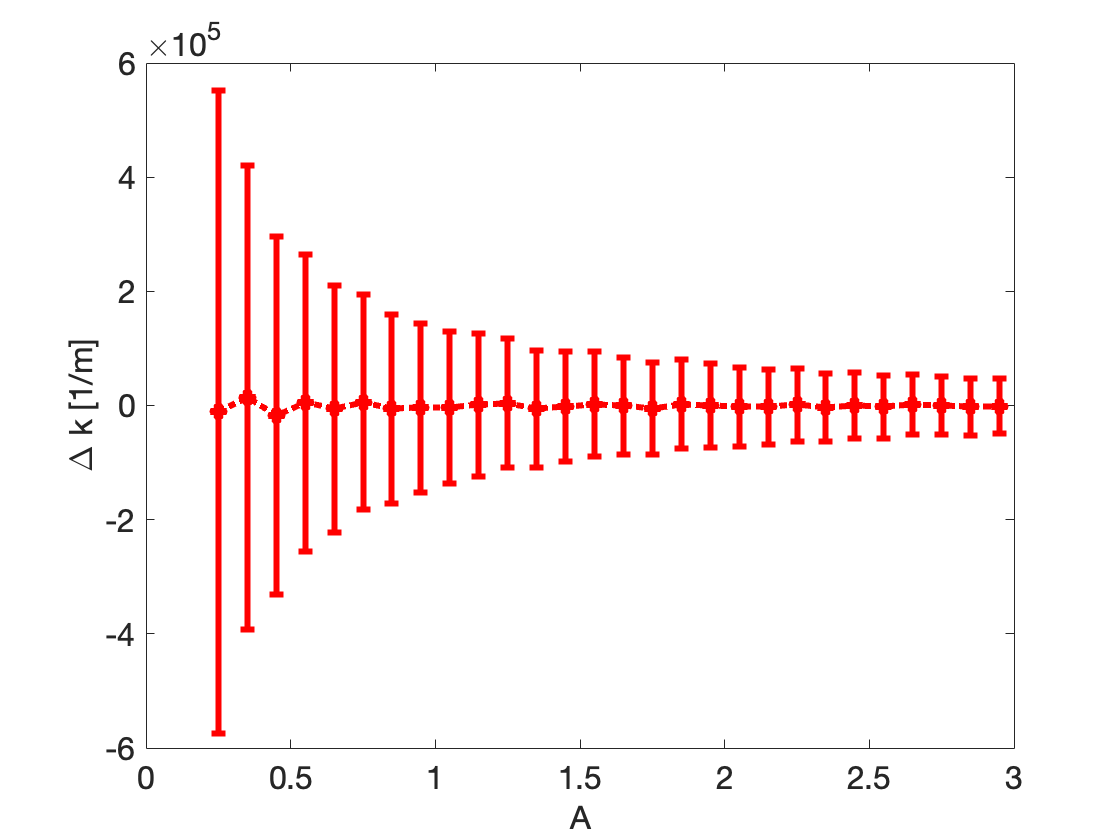} \caption{ Shot-to-shot fluctuations of the MBI-induced central wavenumber shift
$\Delta k$ of the fundamnetal harmonics (n=1) as a function of the normalized seed modulation amplitude $A=\Delta E/E$.
Markers show the mean $\langle \Delta k\rangle$ and error bars indicate the rms
fluctuation $\sigma_{\Delta k}$ obtained from $N=1000$ random-phase realizations
with $\phi_{\mu}\in[0,2\pi)$. The rms jitter decreases with increasing the normalized seed modulation amplitude $A$,
showing that strong seeding mitigates MBI-driven central-frequency fluctuations.}
\label{fig:MBI_flut} 
\end{figure}
This behavior indicates that sufficiently strong seeding not only enhances the
pulse intensity but also mitigates MBI-driven spectral degradation.

The same formalism also allows the evaluation of fluctuations of the central
radiation frequency arising from MBI-driven phase variations. To visualise the shot-to-shot spectral fluctuations induced by MBI, we evaluate
the MBI contribution to the central wavenumber, i.e.\ the last term in
Eq.~(\ref{eq:kvalue}),
\begin{equation}
\Delta k \equiv -\,nB\sum_{\mu} p(k_{\mu})\,k_{\mu}
\cos\!\left(k_{\mu} z + \phi_{\mu}\right).
\end{equation}
For each value of the normalized seed modulation amplitude $A$, we generate an
ensemble of $N=1000$ realizations by sampling the random microbunching phases
$\phi_{\mu}\in[0,2\pi)$ and compute the corresponding distribution of $\Delta k$ for fundamental harmonic (n=1).
Figure~\ref{fig:MBI_flut} reports the mean value $\langle \Delta k\rangle$ and the rms fluctuation
$\sigma_{\Delta k}$ (shown as error bars), demonstrating genuine shot-to-shot
central-frequency jitter driven by MBI. The mean shift remains close to zero,
while the rms jitter decreases with increasing $A$, indicating that stronger
seed-induced modulation suppresses the relative impact of MBI on the spectral
stability.

\section{Conclusion}
The calculated spectral broadening and central-frequency fluctuations
demonstrate that microbunching instability (MBI) not only reduces the
achievable pulse intensity, as discussed above, but also degrades both the
spectral purity and the shot-to-shot spectral stability of the generated THz
radiation in the DALI parameter regime.
These effects become increasingly significant when the MBI-induced energy
modulation becomes comparable to the seed-induced modulation.

In the DALI concept, the electron beam is seeded by radiation extracted from an
FEL oscillator, which inherently limits the available seed power.
As a consequence, operation in a regime of sufficiently strong seeding may not
always be achievable.
Under these conditions, MBI-driven energy modulations can dominate the
microbunching process, leading to a pronounced reduction of pulse intensity,
bandwidth enlargement, and central-frequency jitter.
This highlights a critical operational challenge for DALI and similar
low-energy and high charge THz sources, emphasizing the necessity of careful control and
mitigation of collective effects to ensure stable and spectrally pure THz
output.


\section*{Acknowledgments}

The authors gratefully acknowledge fruitful discussions and support from colleagues at Helmholtz-Zentrum Dresden-Rossendorf (HZDR) and partner institutions. 
In particular, we thank Ulf Lehnert for valuable input throughout this work.We also sincerely appreciate the insightful discussions with Manfred Helm, Sebastian F. Maehrlein, and Stephan Winnerl.

\bibliographystyle{unsrt}
\bibliography{sample}

@inproceedings{Vinokurov1971,
  title        = {The potential of the optical klystron},
  author       = {Vinokurov, N. A. and Skrinsky, A. N.},
  booktitle    = {Proceedings of the 1971 International Conference on High-Energy Accelerators},
  address      = {CERN, Geneva},
  year         = {1971}
}

@article{Coisson1981,
  title        = {Optical klystrons},
  author       = {Coisson, R.},
  journal      = {Particle Accelerators},
  volume       = {11},
  pages        = {245--254},
  year         = {1981},
  url          = {https://cds.cern.ch/record/1107994/files/p245.pdf}
}

@article{Kim1986OK,
  title        = {Analysis of an optical klystron as a high-gain free-electron laser amplifier},
  author       = {Kim, Kwang-Je},
  journal      = {Physical Review Letters},
  volume       = {57},
  number       = {16},
  pages        = {1871--1874},
  year         = {1986},
  doi          = {10.1103/PhysRevLett.57.1871}
}

@article{Yu2000,
  title        = {High-gain harmonic-generation free-electron laser},
  author       = {Yu, L. H. and Ben-Zvi, I. and DiMauro, L. F. and others},
  journal      = {Science},
  volume       = {289},
  pages        = {932--934},
  year         = {2000},
  doi          = {10.1126/science.289.5481.932}
}

@article{Colson1985,
  title        = {The optical klystron free electron laser: theory and experiment},
  author       = {Colson, W. B.},
  journal      = {Nuclear Instruments and Methods in Physics Research A},
  volume       = {237},
  pages        = {1--7},
  year         = {1985},
  doi          = {10.1016/0168-9002(85)90364-5}
}

@article{Minehara1992,
  title        = {Experimental results of the optical klystron free electron laser at UVSOR},
  author       = {Minehara, E. J. and Yamazaki, T. and Kimura, T. and others},
  journal      = {Nuclear Instruments and Methods in Physics Research A},
  volume       = {318},
  number       = {1-3},
  pages        = {131--134},
  year         = {1992},
  doi          = {10.1016/0168-9002(92)90194-L}
}

@article{Takano1991,
  title        = {Performance of the optical klystron FEL at the NIJI-IV storage ring},
  author       = {Takano, S. and Yamazaki, J. and Kimura, T. and others},
  journal      = {Nuclear Instruments and Methods in Physics Research A},
  volume       = {304},
  number       = {1-3},
  pages        = {277--282},
  year         = {1991},
  doi          = {10.1016/0168-9002(91)90214-6}
}

@article{Orzechowski1986,
  title        = {Optical klystron experiment in a free-electron laser amplifier},
  author       = {Orzechowski, T. J. and Duguay, M. A. and Giordano, N. and others},
  journal      = {Physical Review Letters},
  volume       = {57},
  number       = {18},
  pages        = {2172--2175},
  year         = {1986},
  doi          = {10.1103/PhysRevLett.57.2172}
}

@article{Saldin1998,
  author  = {E. L. Saldin and E. A. Schneidmiller and M. V. Yurkov},
  title   = {Optical klystron and self-amplified spontaneous emission FELs},
  journal = {Optics Communications},
  year    = {1998}
}

@book{SaldinBook2000,
  author    = {E. L. Saldin and E. A. Schneidmiller and M. V. Yurkov},
  title     = {The Physics of Free Electron Lasers},
  publisher = {Springer},
  year      = {2000}
}

@book{Dattoli1993,
  author    = {A. Dattoli and A. Renieri and A. Torre},
  title     = {Lectures on the Free Electron Laser Theory},
  publisher = {World Scientific},
  year      = {1993}
}

@article{Williams2006,
  author  = {G. P. Williams},
  title   = {Filling the THz gap -- high power sources and applications},
  journal = {Reports on Progress in Physics},
  year    = {2006}
}

@article{Hemsing2018,
  title = {Bunching phase and constraints on echo enabled harmonic generation},
  author = {Hemsing, E.},
  journal = {Phys. Rev. Accel. Beams},
  volume = {21},
  issue = {5},
  pages = {050702},
  numpages = {8},
  year = {2018},
  month = {May},
  publisher = {American Physical Society},
  doi = {10.1103/PhysRevAccelBeams.21.050702}
}

@article{mirian2021,
  title = {Characterization of soft x-ray echo-enabled harmonic generation free-electron laser pulses in the presence of incoherent electron beam energy modulations},
  author = {Mirian, N. S. and Perosa, G. and others},
  journal = {Phys. Rev. Accel. Beams},
  volume = {24},
  issue = {8},
  pages = {080702},
  numpages = {9},
  year = {2021},
  month = {Aug},
  publisher = {American Physical Society},
  doi = {10.1103/PhysRevAccelBeams.24.080702}
}

@inproceedings{Mirian2025, 
    title = {Concept and preliminary design of the DALI accelerator lattice},
    author = {Mirian, N. and others},
    booktitle = {Proc. IPAC25},
    pages = {114-117},
    paper = {MOPB023},
    venue = {Taipei, Taiwan},
    series = {IPAC25 - 16th International Particle Accelerator Conference},
    number = {16},
    publisher = {JACoW Publishing, Geneva, Switzerland},
    month = {06},
    year = {2025},
    issn = {2673-5490},
    isbn = {978-3-95450-248-6},
    doi = {10.18429/JACoW-IPAC25-MOPB023}
}

@misc{mirian2025ARX,
      title={Short-Pulse High-Power THz Generation Using Optical Klystron FELs: Simulation Results}, 
      author={Najmeh Mirian},
      year={2025},
      eprint={2510.05842},
      archivePrefix={arXiv},
      primaryClass={physics.acc-ph},
      url={https://arxiv.org/abs/2510.05842}, 
}

@article{HuangKim2007,
  author  = {Z. Huang and K.-J. Kim},
  title   = {Review of x-ray free-electron laser theory},
  journal = {Phys. Rev. ST Accel. Beams},
  year    = {2007}
}

@article{Heifets2002,
  author  = {S. Heifets and G. Stupakov and S. Krinsky},
  title   = {Coherent synchrotron radiation instability in a bunch compressor},
  journal = {Phys. Rev. ST Accel. Beams},
  year    = {2002}
}

@article{Saldin2004,
  author  = {E. L. Saldin and E. A. Schneidmiller and M. V. Yurkov},
  title   = {Collective effects in free electron lasers},
  journal = {Nucl. Instrum. Methods Phys. Res. A},
  year    = {2004}
}

@article{SALDIN2004355,
title = {Longitudinal space charge-driven microbunching instability in the TESLA Test Facility linac},
journal = {Nuclear Instruments and Methods in Physics Research Section A: Accelerators, Spectrometers, Detectors and Associated Equipment},
volume = {528},
number = {1},
pages = {355-359},
year = {2004},
issn = {0168-9002},
doi = {https://doi.org/10.1016/j.nima.2004.04.067},
author = {E.L Saldin and E.A Schneidmiller and M.V Yurkov}
}

@article{DiMitri2025,
  title = {Systematic and comprehensive comparison of two semianalytical models of microbunching instability},
  author = {Di Mitri, S. and Campri, G. and others},
  journal = {Phys. Rev. Accel. Beams},
  volume = {28},
  issue = {4},
  pages = {044401},
  numpages = {24},
  year = {2025},
  month = {Apr},
  publisher = {American Physical Society},
  doi = {10.1103/PhysRevAccelBeams.28.044401},
  url = {https://link.aps.org/doi/10.1103/PhysRevAccelBeams.28.044401}
}

@article{Huang2010,
  title = {Measurements of the linac coherent light source laser heater and its impact on the x-ray free-electron laser performance},
  author = {Huang, Z. and Brachmann, A. and others},
  journal = {Phys. Rev. ST Accel. Beams},
  volume = {13},
  issue = {2},
  pages = {020703},
  numpages = {12},
  year = {2010},
  month = {Feb},
  publisher = {American Physical Society},
  doi = {10.1103/PhysRevSTAB.13.020703},
  url = {https://link.aps.org/doi/10.1103/PhysRevSTAB.13.020703}
}

@article{Huang2004,
  title = {Suppression of microbunching instability in the linac coherent light source},
  author = {Huang, Z. and Borland, M. and others},
  journal = {Phys. Rev. ST Accel. Beams},
  volume = {7},
  issue = {7},
  pages = {074401},
  numpages = {10},
  year = {2004},
  month = {Jul},
  publisher = {American Physical Society},
  doi = {10.1103/PhysRevSTAB.7.074401},
  url = {https://link.aps.org/doi/10.1103/PhysRevSTAB.7.074401}
}

@Article{Perosa2021,
author={Perosa, Giovanni
and Di Mitri, Simone},
title={Matrix model for collective phenomena in electron beams longitudinal phase space},
journal={Scientific Reports},
year={2021},
month={Apr},
day={12},
volume={11},
number={1},
pages={7895},
issn={2045-2322},
doi={10.1038/s41598-021-87041-0},
url={https://doi.org/10.1038/s41598-021-87041-0}
}

@article{DiMitri2020,
  title = {Experimental evidence of intrabeam scattering in a free-electron laser driver},
  author = { Di Mitri, S. and  Perosa, G. and others},
  journal = {New J. Phys},
  volume = {22},
  pages = { 083053},
  year = {2020},
  month = {August},
}

@article{Huang2002,
  title = {Formulas for coherent synchrotron radiation microbunching in a bunch compressor chicane},
  author = {Huang, Z. and Kim, K. -J.},
  journal = {Phys. Rev. ST Accel. Beams},
  volume = {5},
  issue = {7},
  pages = {074401},
  numpages = {7},
  year = {2002},
  month = {Jul},
  publisher = {American Physical Society},
  doi = {10.1103/PhysRevSTAB.5.074401},
  url = {https://link.aps.org/doi/10.1103/PhysRevSTAB.5.074401}
}

@article{Stupakov2002,
  title = {Coherent synchrotron radiation instability in a bunch compressor},
  author = {Heifets, S. and Stupakov, G. and Krinsky, S.},
  journal = {Phys. Rev. ST Accel. Beams},
  volume = {5},
  issue = {6},
  pages = {064401},
  numpages = {10},
  year = {2002},
  month = {Jun},
  publisher = {American Physical Society},
  doi = {10.1103/PhysRevSTAB.5.064401},
  url = {https://link.aps.org/doi/10.1103/PhysRevSTAB.5.064401}
}

@report{DALI-CDR,
 author       = {Helm, M. and others}, 
	title        ={DALI Conceptual Design Report}, 
	 year         = {2024},
	url={https://www.hzdr.de/DALI}
	}

@report{DALI-TDR,
 title={DALI Internal report},
 author       = {Helm, M. and others}, 
	 year         = {2025},
	url={https://www.hzdr.de/DALI}
	}

@article{Penco2017,
  author  = {Penco, G. and others},
  title   = {Optical Klystron Enhancement to Self Amplified Spontaneous Emission at FERMI},
  journal = {Photonics},
  year    = {2017},
  volume  = {4},
  number  = {1},
  pages   = {15},
  doi     = {10.3390/photonics4010015},
}

@article{Penco2015,
  title = {Experimental Demonstration of Enhanced Self-Amplified Spontaneous Emission by an Optical Klystron},
  author = {Penco, G. and Allaria, E. and others},
  journal = {Phys. Rev. Lett.},
  volume = {114},
  issue = {1},
  pages = {013901},
  numpages = {5},
  year = {2015},
  month = {Jan},
  publisher = {American Physical Society},
  doi = {10.1103/PhysRevLett.114.013901},
  url = {https://link.aps.org/doi/10.1103/PhysRevLett.114.013901}
}

@article{Ding2006,
  title = {Optical klystron enhancement to self-amplified spontaneous emission free electron lasers},
  author = {Ding, Y. and Emma, P. and Huang, Z. and Kumar, V.},
  journal = {Phys. Rev. ST Accel. Beams},
  volume = {9},
  issue = {7},
  pages = {070702},
  numpages = {7},
  year = {2006},
  month = {Jul},
  publisher = {American Physical Society},
  doi = {10.1103/PhysRevSTAB.9.070702},
  url = {https://link.aps.org/doi/10.1103/PhysRevSTAB.9.070702}
}

@article{Zhang2017,
  title = {Generation of high-power, tunable terahertz radiation from laser interaction with a relativistic electron beam},
  author = {Zhang, Z. and Yan, L. and others},
  journal = {Phys. Rev. Accel. Beams},
  volume = {20},
  issue = {5},
  pages = {050701},
  numpages = {9},
  year = {2017},
  month = {May},
  publisher = {American Physical Society},
  doi = {10.1103/PhysRevAccelBeams.20.050701},
  url = {https://link.aps.org/doi/10.1103/PhysRevAccelBeams.20.050701}
}

@article{Thomas2010,
  title = {Algorithm for calculating spectral intensity due to charged particles in arbitrary motion},
  author = {Thomas, A. G. R.},
  journal = {Phys. Rev. ST Accel. Beams},
  volume = {13},
  issue = {2},
  pages = {020702},
  numpages = {11},
  year = {2010},
  month = {Feb},
  publisher = {American Physical Society},
  doi = {10.1103/PhysRevSTAB.13.020702},
  url = {https://link.aps.org/doi/10.1103/PhysRevSTAB.13.020702}
}

\end{document}